\documentclass[amsmath,amssymb,onecolumn,superscriptaddress]{revtex4}
\usepackage{graphicx}% Include figure files
\usepackage{dcolumn}% Align table columns on decimal point

\def\be{\begin{equation}}
\def\ee{\end{equation}}
\def\beq{\begin{eqnarray}}
\def\eeq{\end{eqnarray}}

\usepackage{bm}% bold math
\usepackage{graphicx}
\usepackage{dcolumn}
\usepackage{amsmath}
\usepackage[latin1]{inputenc}
\usepackage{graphicx, psfrag}
\usepackage{amssymb}
\usepackage[colorlinks=true, citecolor=blue, urlcolor = blue, linkcolor= red, bookmarks=true]{hyperref}
\usepackage{float}
\usepackage{amsmath}
\usepackage{tensor}
\usepackage{amsfonts}
\usepackage{dcolumn}
\usepackage{hyperref}
\usepackage{subfigure}
\usepackage{pgfplots}
\usepackage{epstopdf}
\usepackage{booktabs}
\usepackage{pdfpages}

\begin{document}
\title{Deflection angle evolution with plasma
medium and without plasma medium in a parameterized black hole}

\author{Xiaoling He}
\email{hxlfudan@aliyun.com}
\affiliation{School of Science, Zhejiang University of Science and Technology, Hangzhou 310023, Zhejiang, China.}

\author{Tianyu Xu}
\email{744549568@qq.com}
\affiliation{School of Science, Zhejiang University of Science and Technology, Hangzhou 310023, Zhejiang, China.}

\author{Yun Yu}
\email{1957517787@qq.com}
\affiliation{Zhejiang University of Science and Technology, Hangzhou 310023, Zhejiang, China}

\author{Anosha Karamat}
\email{anosha.karamat2@gmail.com}
\affiliation{Department of Mathematics, Minhaj University, Lahore-54000, Pakistan}

\author{Rimsha Babar}
\email{rimsha.babar10@gmail.com}
\affiliation{Division of Science and Technology, University of Education, Township, Lahore-54590, Pakistan}

\author{Riasat Ali}
\email{riasatyasin@gmail.com}
\affiliation{Department of Mathematics, GC
University Faisalabad Layyah Campus, Layyah-31200, Pakistan}

\begin{abstract}
Using the Keeton and Petters approach, we determine the deflection angle. We also investigate the motion of photons around a
parameterized black hole in the presence of non-magnetized cold plasma by using a new ray-tracing algorithm. In spherically
symmetric spacetime, we examine the influence of the plasma by applying the Hamiltonian
equation on the deflection angle as well as shadow. It is examine to derive the rays analytically from
Hamilton's equation by separating the metric and the plasma frequency. We study that the presence of
plasma affects the deflection angle as well as shadow for the parameterized black hole and they depend on plasma frequency.
If the plasma frequency is significantly lower than the photon frequency, the photon sphere and shadow radius expressions can be linearized around the values found for space light rays.
Furthermore, we have graphically analyze the behavior of shadow for distinct positions under the effects of plasma frequency as well as low density plasma medium.
\end{abstract}

\date{\today}

\maketitle

\section{introduction}

Black holes are invisible objects that are typically thought to arise in the
gravitational collapse of enormous astronomical objects, as predicted by
Einstein's general relativity (GR). It is commonly known that photons
released from a lit source behind a black hole (BH) will create the so-called
BH shadow, a two-dimensional dark region in the observer's sky. The BHs shadow,
which is an impression of the BH, tells us important things about the BH.
For instance, one can determine the BHs charge and spin \cite{1,2,3,4,5,6}
as well as limit certain additional parameters brought about by changing gravities
\cite{7,8,9,10,11} from the shadow of the BH. First, Synge \cite{12} and later
Luminet \cite{13} looked at the perfect circle-shaped shadow that a spherically
symmetric BH casts. They also presented formulas to determine the shadow size
and angular radius, respectively. Bardeen \cite{14} was the first to examine the
shadow produced by a Kerr BH. Because of the dragging action, the shadow shape
is modified. Since then, there has been a lot of research done in the literature
on the BH shadow. For instance, \cite{15} looked at the BH shadow and photon sphere
in dynamically changing spacetimes.

The gravitational lensing effect is the process by which BHs bend the escaping
light beams in addition to capturing them. Gravitational lensing is another
effective approach that gives us a vast amount of data on BHs, including their
location, angular momentum and mass. The geodesic technique \cite{16,17,18,19,20}
has been intensively researched for gravitational lensing for black holes, wormholes,
cosmic strings and other phenomena since the first measurement of the deflection
angle of light by the sun. Gibbons and Werner presented a different approach in
ref. \cite{21} for calculating the weak deflection angle of light by a spherically
symmetric BH within the framework of optical geometry using the Gauss-Bonnet theorem and then,
Werner \cite{22} extended this method to stationary BHs by applying Kerr-Randers optical metric.Â 
Additionally, whereas the impact of plasma on light rays can often be disregarded in
astronomical contexts, this is not the case for light rays in the radio frequency spectrum.
The effects of the solar corona, which is thought of as a non-magnetized pressureless plasma,
on the time delay \cite{23} and deflection angle when the light rays propagate close to the
Sun \cite{24} are one well known examples. Later, Perlick studied the impact of a non-homogeneous
plasma on aÂ light deflection in the equatorial plane of the Kerr metric and in the Schwarzschild
metric \cite{25}. Since then, research on how plasma affects light transmission has become
increasingly popular. For instance, refs. \cite{26,27,28,29,30,31} looked at the effects of
plasma on gravitational lensing by BHs and compact objects. In refs. \cite{32,33,34,35},
the shadows of wormholes and BHs surrounded by plasma were studied.

One of the most important reasons to study quantum gravity is the search for ultraviolet complete
gravitational theories that could prevent space-time singularity. There are numerous attempts,
including string field theory and quantum gravity, even though we do not yet have a fully formed
quantum gravity theory. According to a common assumption shared by nearly all of these theories,
there should be an intrinsic modified geometry in space-time \cite{36} that is on the order of the
Planck length. A configuration with this much extension suggests that space-time is not local
\cite{37,38,39,40}. So, it is expected that the linear correction to GR is the first-order
correction \cite{41,42,43,44,45,46,47,48, z14c} of quantum theory. The quantum gravity and shadow
are compulsory components to study BH geometry.

The deflection angle for Einstein-Maxwell-Dilaton-Axion like BH using the methodology developed by Gibbons and Werner has been studied \cite{RA1}. Further evidence that the gravitational lensing effect is a global and even topological phenomenon, i.e., there are several light rays converging between the source and observer, was provided by the fact that the deflection of a light ray was determined outside of the lensing area. The weak limit approach \cite{RA2} of the Gauss-Bonnet theorem was used to compute the deflection angle of a null aether like BH by investigating the light beams in a BH gravitational field. Matsuno calculated \cite{RA3} the deflection angles while taking into account photon movements around a charged squashed Kaluza-Klein like BH particle in an unmagnetized homogeneous cold plasma medium. The light deflection angle at the leading order terms was determined using the Gauss-Bonnet theorem and a straight line approach by \cite{RA4} and also examined the gravitational lensing of the Einstein-Cartan-Kibble-Sciama like BH in the weak field approximation. The Gauss-Bonnet theorem has been used \cite{RA5} to study weak gravitational lensing in Kerr-Newman-Kasuya like optical spacetime, where optical geometry provides a more geometrical view. It was found that the deflection angle decreased linearly with the additional magnetic charge compared to Kerr-Newman BHs.

In our work, we explore the deflection angle by using the Keeton and Petters technique and ray-tracing algorithm from the photon
pathway around a BH in the presence of non-plasma and plasma with radial density and evaluate the impacts of plasma
on the BH shadow.
We have formulated our paper by the following pattern: In Sec. \textbf{II}, we review the BH metric and study the effects of deformation parameter to metric function. In Sec. \textbf{III}, using the Keeton and Petters approach, to determine the deflection angle. In Sec. \textbf{IV} explains the motion equation
of light ray in the scenario of BH with plasma. In Sec. \textbf{V}, we
study the circular light orbits that are very important for the creation of the shadow.
Sec. \textbf{VI}, computes the angular radius of the BH shadow.
In Sec. \textbf{VII}, we graphically discuss the impacts of different parameters on BH Shadow.
Sec. \textbf{VIII}, examines the case of a low-density plasma Shadow.
In Sec. \textbf{IX}, we summarize the results.

\section{Deflection angle analysis with plasma medium}

Huang et al. used \cite{DA} a new ray-tracing algorithm to examine the photon's velocity around a BH in the presence of plasma, whose density is a function of the radius coordinate, and look into the impact of the plasma on the BHs shadow. We expect that the spacetime is loaded up with a non-magnetized cold plasma whose electron plasma frequency $w_p$ only seems on the coordinate of radius ($r$) in the form
\begin{equation}
w^{2}_{p}(r)=\frac{4\pi E(r) e^{2}}{m},\label{b3}
\end{equation}
whereas $E(r)$, $m$ and $e$ are the number of electron's density in the non-magnetized plasma, electron mass and electron charge, respectively.
This plasma's refraction index $i$, is influenced by the photon's frequency as measured by a static observer and the radius coordinate $r$ in the following way
\begin{equation}
i^{2}(r,w)=1-\frac{w_{p}^{2}(r)}{w^{2}}.\label{b4}
\end{equation}

Due to the spherical symmetry, we can limit to the
equatorial plane $\theta=\frac{\pi}{2}, ~u_{\theta}=0$. The Hamiltonian explains the particle's motion in a pressure-free and non-magnetized plasma as
\begin{eqnarray}
H&=&\frac{1}{2}\left(g^{ik}u_{i}u_{k}+w^{2}_{p}(r)\right),\nonumber\\
&=&\frac{1}{2}\left(-\frac{u^{2}_{t}}{A(r)}+\frac{u^{2}_{r}}{B(r)}+\frac{u^{2}_{\phi}}{C(r)}+w^{2}_{p}(r)\right).\label{b5}
\end{eqnarray}
Light waves for Hamilton equations imply that
\begin{equation}
\dot{u}_{i}=-\frac{\partial H}{\partial \delta^{i}},~~~~\dot{\delta}^{i}=\frac{\partial H}{\partial{u}_{i}},\label{b6}
\end{equation}
it can be written as
\begin{eqnarray}
\dot{u}_{t}&=&-\frac{\partial H}{\partial t}=0,\label{b7}\\
\dot{u}_{\phi}&=&-\frac{\partial H}{\partial \phi}=0,\label{b8}\\
\dot{u}_{r}&=&-\frac{\partial H}{\partial r}
=\frac{1}{2}\left(-\frac{u^{2}_{t}\acute{A}(r)}{A^{2}(r)}+
\frac{u^{2}_{r}\acute{B}(r)}{B^{2}(r)}+\frac{s^{2}_{\phi}\acute{C}(r)}{C^{2}(r)}-\frac{d}{dr}{w^{2}}_{p}(r)\right),\label{b9}\\
\dot{t}&=&\frac{\partial H}{\partial u_{t}}=-\frac{u_{t}}{A(r)},\label{b10}\\
\dot{r}&=&\frac{\partial H}{\partial u_{r}}=\frac{u_{r}}{B(r)},\label{b10a}\\
\dot{\phi}&=&\frac{\partial H}{\partial u_{\phi}}=\frac{u_{\phi}}{C(r)},\label{b11}
\end{eqnarray}
for $H=0$, we have
\begin{equation}
0=-\frac{u^{2}_{t}}{A(r)}+\frac{u^{2}_{r}}{B(r)}+\frac{u^{2}_\phi}{C(r)}+{w^{2}_{p}(r)},\label{b12}
\end{equation}
here, dot and prime shows differentiation w. r. t an affine
parameter and $r$, respectively.
It follows from Eq. (\ref{b7}) and (\ref{b8}) that $u_{t}$ and $u_{\phi}$ represents constants of motion. We may type $w_{0}=-u_{t}$. In case of a asymptotically flat spacetime, the $w_{0}$ is fixed and $A(r)\rightarrow 1$ as $r\rightarrow{\infty}$ and the frequency of a static observer represented by $w(r)$ can be defined by redshift formula as
\begin{equation}
w(r)=\frac{w_{0}}{\sqrt{A(r)}}.\label{b14}
\end{equation}
From the equation (\ref{b12}), the constant motion $w_{0}$ of light beam is specified to the particular region with
\begin{equation}
\frac{w^{2}_{0}}{A(r)}>{w^{2}_{p}(r)}.\label{b15}
\end{equation}
The photons frequency at that place should be greater than the plasma frequency, according to Eq. (\ref{b15}).
It is true for the light transmission in a medium of plasma.
We consider Eqs. (\ref{b10a}) and (\ref{b11}), to get the equation of orbit
\begin{equation}
\frac{dr}{d\phi}=\frac{\dot{r}}{\dot{\phi}}=\frac{C(r){u}_{r}}{B(r){u}_{\phi}}.
\end{equation}
By substituting for $u_{r}$ from equation (\ref{b12}), which is taken by
\begin{equation}
\frac{dr}{d\phi}=\pm\frac{\sqrt{C(r)}}{\sqrt{B(r)}}\left(\sqrt{\frac{w^{2}_{0}}{u^{2}_{\phi}}{D(r)^{2}}-1}\right),\label{b17}
\end{equation}
with
\begin{equation}
\frac{w^{2}_{0}}{u^{2}_{\phi}}{D(r)^{2}}-1>0,~~
{D^{2}(r)}=\frac{C(r)}{A(r)}\left({1}-{A(r)}\frac{w^{2}_{p}(r)}{w^{2}_{0}}\right).\label{b18}
\end{equation}
In this case, the equation (\ref{b17}) sign-in must be suitably followed and the orbit should be divided
into segments at the points $r$ is slowly decreasing and increasing in connection to $\phi$. The expression for the
deflection angle ($\Psi$) for an infinite light beam with a radius of $Y$ and an additional infinity is
produced by integrating over the orbit, we get
\begin{equation}
\pi+\Psi={2}\int_{Y}^{\infty}\frac{\sqrt{{B}(r)}}{\sqrt{C(r)}}\left(\frac{w^{2}_{0}}{u^{2}_{\phi}}{D^{2}(r)}-{1}\right)^{-\frac{1}{2}}{dr}.\label{b19}
\end{equation}
Since $Y$ is related with the pivot of trajectory, the specified case $\frac{dr}{d\phi}|_{Y}=0$
must be fulfilled. The above mentioned equation determines the following relationship between the
radius $Y$ and the constant motion $\frac{u_{\phi}}{w_{0}}$ as follows
\begin{equation}
{D^{2}(X)}=\frac{u^{2}_{\phi}}{w^{2}_{0}}.\label{b20}
\end{equation}
In terms of $w$ and $Y$, the angle of deflection for a certain plasma distribution can be expressed as follows
\begin{equation}
\pi+\Psi=2\int_{Y}^{\infty}\left(\frac{\frac{r^5}{r^{3}-\eta-2Mr^{2}}-\frac{r^{2}w^{2}_{p}(r)}{w^{2}_{0}}}
{\frac{Y^{5}}{Y^{3}-\eta-2MY^{2}}-\frac{Y^{2}w^{2}_{p}(Y)}{w^{2}_{0}}}-1\right)^{-\frac{1}{2}}
\frac{dr}{\sqrt{r^{3}-\eta-2Mr^{2}}}.\label{b24}
\end{equation}
The deflection angle is dependent by the horizon radius, the infinite radius of the light wave $Y$, the BH mass,
the photon and plasma frequency, the BH geometry, and BH deformation parameter.
When $\eta=0$, the equation (\ref{b24}) for the plasma's deflection angle on Schwarzschild spacetime was developed in \cite{DAN1}. In \cite{DAN2}, it was derived using Synge's method and formulated in terms of an elliptic integral for a homogeneous plasma. Further, the strong deflection limit $\Psi\gg 1$ was examined. Moreover, we conclude that for $\eta=0$, we recover the deflection angle for Schwarzschild BH as calculated in \cite{33}.

\section{Orbits of Light}
For photons emitted from the observer's screen in the previous ray-tracing method, there are
only two possible final states: either the photon has been absorbed by the BH or it is scattered
indefinitely. Once released, the photons will gradually scatter around the BH, which is where the
shadow of a BH is situated in the observer's image plane. This section will discuss the condition
of circular light orbits, which is necessary for locating the shadow. It is important to note that
the light is flowing in a circular direction while the states $\dot{r}=0$ and $\ddot{r}=0$.
Eqs. (\ref{b10a}) and (\ref{b12}) with $u_r=0$ give us the relation shown below:
\begin{equation}
0=-\frac{w^{2}_{0}}{A(r)}+\frac{u^{2}_{\phi}}{C(r)}+{w^{2}_{p}(r)}.\label{b27}
\end{equation}
As an approach, (\ref{b10a}) implies
\begin{equation}
\dot{u}_{r}=\frac{d}{d\lambda}(B(r)\dot{r})=\ddot{r}B(r)+\dot{r}^2 {B^\prime}(r).\label{b28}
\end{equation}
If $\dot{r}=0$, $\ddot{r}=0$ and $\dot{u}_r=0$, then Eq. (\ref{b8}) represents the light orbits second equation
\begin{equation}
0=-\frac{w^{2}_{0}A^\prime(r)}{A^2(r)}+\frac{u^{2}_{\phi}C^\prime(r)}{C^2(r)}-\frac{d}{dr}{w^{2}_{p}(r)}.\label{b29}
\end{equation}
For $u^{2}_{\phi}$, we canÂ compute the solution from Eqs. (\ref{b27}) and (\ref{b29}) as
\begin{eqnarray}
u^2_{\phi}&=&C(r)\left(\frac{w^{2}_{0}}{A(r)}-{w^{2}_{p}(r)}\right),\\
u^2_{\phi}&=&\frac{C^{2}(r)}{C^\prime (r)}\left(\frac{w^{2}_{0}A^\prime(r)}{A^2(r)}+\frac{d}{dr}{w^{2}_{p}(r)}\right).
\end{eqnarray}
By eliminating these two equations from one another after making a few straightforward adjustments provides
the following result for the radius of a circular light orbit in the form
\begin{equation}
0=r^{2}\Big(5r^{3}-5\eta-10r^{2}M-3r^{3}+4r^{6}M\Big)-\Big(r^{3}-\eta-2r^{2}M\Big)^{2}\Big[\frac{2w^{2}_{p}(r)}{w^{2}_{0}}-\frac{r^{2}}{w^{2}_{0}}\frac{d}{dr}{w^{2}_{p}(r)}\Big],\label{b31}
\end{equation}
at which the property $D^{2}(r)$ is produced by the formula (\ref{b18}). The photon sphereÂ 
radiusÂ can be calculated by using the (\ref{b31}) solution for $r=r_p$. A light moving
tangentially into a sphere always goes in a circle with radius $r=r_p$.
The furthest photon circle is consistently unstable in terms of radial disorder if the BH space-time is
asymptotically flat with $w_{p}\rightarrow A 0$ and $r\rightarrow\infty$. This implies that the asymptotic
light beams that reach them could act as limit curves for the circular photon's orbits. The exterior
photon circle's radius and the analytical point of the recently discussed un-important radius $X$ are identical.
A light ray goes to its point of origin when it leaves its infinity-based origin and goes to a minimum
radius $Y$ that is higher than $r_p$. A light beam spirals asymptotically along the direction of a photon
sphere's circular orbit in the case where $Y=r_p$. The fact that the equation (\ref{b31}) can be reduced to
the Atkinson scenario \cite{49} for $w_p(r)=0$ is crucial.
\section{SHADOW RADIUS}

The shadow is a visualization of a spherically symmetric space-time. By taking into account the description of the expression from the preceding section, we examine the effects of plasma on an incident light. We discuss the two classifications of these incident photons. Through the BH, photons of first-class light are reflected back and move continuously. The photon of second-class light is going toward the BH's horizon.
\begin{center}
\includegraphics[width=10cm]{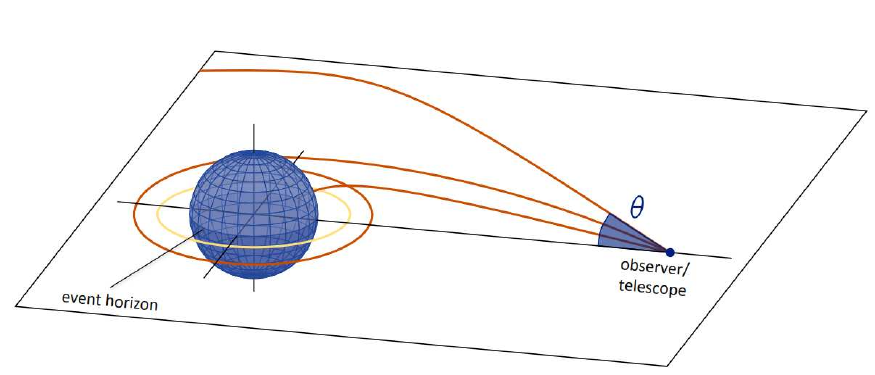}
\end{center}

Let's start by supposing that there isn't a source of light in the observer's sky that would interfere with the BH-observer relationship. The BH's shadow is represented by the sphere in the observer's sky. The starting trajectory of the photon's light, which is asymptotically radial in the direction of the outer photon sphere, determines the shadow radius.
\begin{equation}
\cot\Phi=\frac{g_{rr}}{g_{\phi\phi}}\frac{dr}{d\phi}|_{r=r_{0}}=\frac{\sqrt{B(r)}}{\sqrt{C(r)}}\frac{dr}{d\phi}|_{r=r_{0}}.
\end{equation}
Using the equation (\ref{b20}) to the orbit equation (\ref{b17}), we obtain the minimumÂ ray of light as it approaches a minimal radius Y.
\begin{equation}
\frac{dr}{d\phi}=\pm\frac{\sqrt{C(r)}}{\sqrt{B(r)}}\sqrt{\frac{D^2(r)}{D^2(X)}-1}.
\end{equation}
For the angle $\Phi$, we obtain the following:
\begin{equation}
\cot^2\Phi=\frac{D^2(r_{0})}{D^2(X)}-1.
\end{equation}
This suggests
\begin{equation}
\sin^2\Phi=\frac{D^2(r_{0})}{D^2(X)}.\label{b34}
\end{equation}
The limit of shadows is determined by the approximate rotation of light into a sphere with a radius of $r_p$. As a result, sending $X\rightarrow r_{p}$ gives the angular shadow size in Eq. (\ref{b34}) as
\begin{equation}
\sin^2\Phi=\frac{D^2(r_{ph})}{D^2(r_{0})},\label{b35}
\end{equation}
where the equation expression (\ref{b18}) expresses $D(r)$.
We can assume that the observer is situated in a zone of low-density plasma.
The  equation result (\ref{b18}) gives
\begin{equation}
D^2(r_{0})=\frac{C(r_{0})}{A(r_{0})},\label{b36}
\end{equation}
and equation (\ref{b35}) implies
\begin{equation}
\sin^2\Phi=\frac{r^{5}_{ph}(r^{3}_{0}-\eta-2Mr^{2}_{0})}{r^{5}_{0}(r^{3}_{ph}-\eta-2Mr^{2}_{ph})}
\left[1-\Big(1-\frac{\eta}{r^{3}_{ph}}-\frac{2M}{r_{ph}}\Big)\frac{w_{p}^2(r_{ph})}{w^2_{0}}\right].\label{b37}
\end{equation}
This demonstrates how the plasma is gradually affecting the shadow's decreasing radius.
It seems to be the case for all spherically symmetric spacetimes, such as wormhole metrics and compact metrics with an instability photon sphere.
These objects would cast shadows like BHs if the light were not shining from the direction of the centre object towards the observer.
In order for the light rays in the photon sphere to function as limit curves, it is absolutely necessary for them to be unstable in relation to radial perturbations. Because of this, the shadow can be constructed in any static, spherically symmetric spacetime that accepts an unstable photon sphere. This includes black holes as well as wormholes.
The outermost photon sphere is always unstable in a static, asymptotically flat, spherically symmetric spacetime, provided that the plasma density tends to zero for $r\rightarrow\infty$.
It is also worth mentioning here that for $\eta=0$ and $r_{ph}=3M$ in Eq. (\ref{b37}), we recover the radius of shadow for Schwarzschild BH \cite{33}.

\section{Graphical Analysis for Shadow of Black Hole}
This section gives the graphical behavior for shadow of parameterized BH.

\begin{center}
\includegraphics[width=7cm]{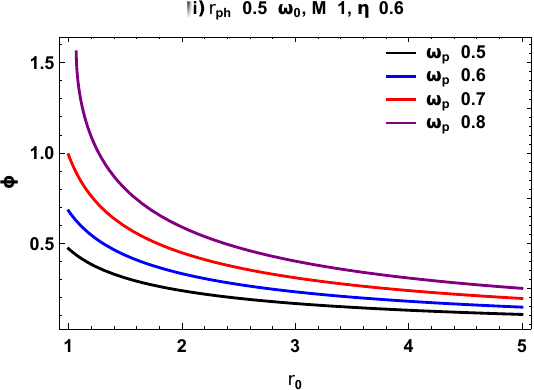}\includegraphics[width=7cm]{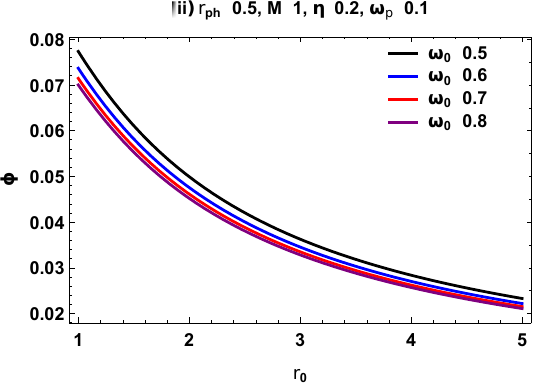}\\
{Figure 2: The BH shadow $\Phi$ versus  radius $r_{0}$ for various $\omega_{p}$, $\omega_{0}$ and fixed $r_{ph}=0.5=\omega_{0}, M=1~\text{and}~ \eta=0.6$ in the left plot while for fixed $r_{ph}=0.5, \omega_{0}=0.1, M=1~\text{and}~\eta=0.2$ in the right plot.}
\end{center}
The left plot of {\bf Fig. 2} shows the behavior of the angular radius of shadow with static observer at distinct positions $r_{0}$ for varying $\omega_{p}$. The $\Phi$ slowly decreases and then, it gets a flat form for rising $r_{0}$. It is notable that the $\Phi$ rises for rising values of $\omega_{p}$.

In the right plot of {\bf Fig. 2}, the $\Phi$ represents the behavior with $r_{0}$ for changing values of $\omega_{0}$. The $\Phi$ slowly decreases till $r_{0}\rightarrow \infty$. The $\Phi$ decreases with rising $\omega_{0}$.

From {\bf Fig. 2}, we conclude that the BH shadow is smaller for different variations of $r_{0}$. To get large shadow radius in strong gravity, the frequency parameter must be extremely small.

\begin{center}
\includegraphics[width=7cm]{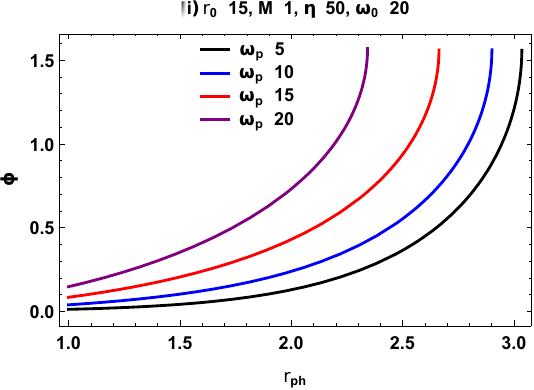}\includegraphics[width=7cm]{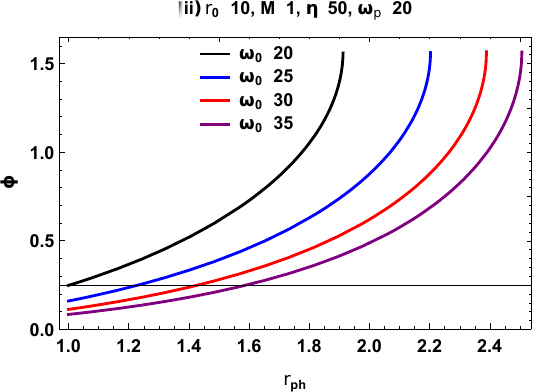}\\
{Figure 3: The BH shadow angular radius w.r.t $r_{ph}$ for various $\omega_{p}$, $\omega_{0}$ and fixed $M=1, \eta=50, \omega_{0}=20$.}
\end{center}
The left plot in {\bf Fig. 3} represents the dependency of $\Phi$ versus $r_{ph}$ for different values of $\omega_{p}$. In the left plot, the $\Phi$ continuously goes on increasing. The $\Phi$ reduces for rising values of $\omega_{p}$.
In the right plot of {\bf Fig. 3}, the $\Phi$  shows a rising attitude for rising $r_{ph}$ as well as the radius of shadow increase for rising values of $\omega_{0}$.

The {\bf Fig. 3} demonstrates that the BH shadow is smaller for different variations of $r_{ph}$ but it shows an increasing behavior. To get large shadow radius in strong gravity, the frequency parameter must be extremely small.
\section{A Low Density Plasma's Shadow}
If the plasma frequency is significantly lower than the photon frequency, the photon sphere and shadow radius expressions can be linearized around the values found for space light rays.
This is addressed by modifying equation (\ref{b18}) as
\begin{equation}
D^2(r)=Z(r)(1-\varsigma \alpha(r)),\label{b38}
\end{equation}
whereas
\begin{equation}
Z(r)=\frac{C(r)}{A(r)},~~ \alpha(r)=\frac{A(r)w^2_{p}(r)}{w^2_{0}}=\frac{w^2_{p}(r)}{w^2(r)}.\label{b39}
\end{equation}
We include an accounts parameter $\varsigma$, which is set to unity once all results have been linearized.Â 
This consequence of linearity can be deduced by power law i.e., $\frac{w^2_{p}(r)}{w^2(r)}=\beta_{0}\frac{M^{k}}{r^{k}}$, where $\beta_{0}>0$ and $k\geq0$ that represents dimensionless constants. A detailed example for Schwarzschild black hole for this consequence is given in \cite{[A1]}.
Now, the photon sphere result (\ref{b31}) written as
\begin{equation}
0=Z^\prime(r)(1-\varsigma \alpha(r))-\varsigma Z(r) \alpha^{\prime}(r).\label{b40}
\end{equation}
It is possible to express the equation's solution as follows
\begin{equation}
r_{ph}=r^{0}_{ph}+\varsigma r^{1}_{ph}+...,\label{b41}
\end{equation}
in the absence of plasma with a solution $r^{0}_{ph}$ such that
\begin{equation}
Z^\prime(r^{0}_{ph})=0.\label{b42}
\end{equation}
We derive the $\varsigma$ coefficients by inserting these formulas into (\ref{b40}) and after comparing them, we get
\begin{equation}
r^{1}_{ph}=\frac{Z^(r^{0}_{ph})\alpha^\prime (r^{0}_{ph})}{Z''(r^{0}_{ph})}.\label{b43}
\end{equation}
In relation to the values of $\alpha^\prime (r^{0}_{ph})$ and $Z''(r^{0}_{ph})$, this equation can be either positive or negative, so that the plasma can move the light sphere outside or inside. Now, we implement an expansion (\ref{b41}) to the shadow result (\ref{b35}). Once more set all components equal to unity while ignoring any quadratic and higher order factors, we obtain
\begin{equation}
\sin^2\Phi=\frac{\tilde{r}^{5}(r_{0}^{3}-\eta -2r_{0}^{2}M)}{r_{0}^{5}\big(\tilde{r}^{3}-\eta -2\tilde{r}^{2}M\big)}\left[1-\Big(1-\frac{\eta}{\tilde{r}^{3}}-\frac{2M}{\tilde{r}}\Big)\frac{w^{2}_{p}(\tilde{r})}{w_{0}^{2}}+\Big(1-\frac{\eta}{r_{0}^{3}} -\frac{2M}{r_{0}}\Big)\frac{w^{2}_{p}(r_{0})}{w_{0}^{2}}\right],\label{b44}
\end{equation}
with $\tilde{r} =r^{0}_{ph}$. Note that the computation for to within this order no longer takes $r_{ph}^{1}$ into account.
From the Eq. (\ref{b44}), the plasma has less of an impact on the shadow than $(1-\frac{\eta}{r_{0}^{3}} -\frac{2M}{r_{0}})\frac{w^{2}_{p}(r_{0})}{w_{0}^{2}}<(1-\frac{\eta}{\tilde{r}^{3}}-\frac{2M}{\tilde{r}})\frac{w^{2}_{p}(\tilde{r})}{w_{0}^{2}}$.
The photon sphere radius $r_{ph}$ must be computed for a given metric (\ref{sol}), a defined photon frequency at infinity $w_{0}$, and a specified plasma frequency that satisfies the case $w_{p}(r)<<w(r)$. Approximate
$r^{0}_{ph}$ from (\ref{b42}) and $r^{1}_{ph}$ from (\ref{b43}), then estimate $r_{ph}$
allowing to (\ref{b43}) with $\varsigma=1$. Then, write down the functions $Z(r)$ and
 $\alpha(r)$ (see (\ref{b39})). In this procedure, we have to put $r^{0}_{ph}$ into the expression to obtain the
 radius of shadow angular $\Phi$ for observer particular $r_{0}$ position (\ref{b44}).

\begin{center}
\includegraphics[width=7cm]{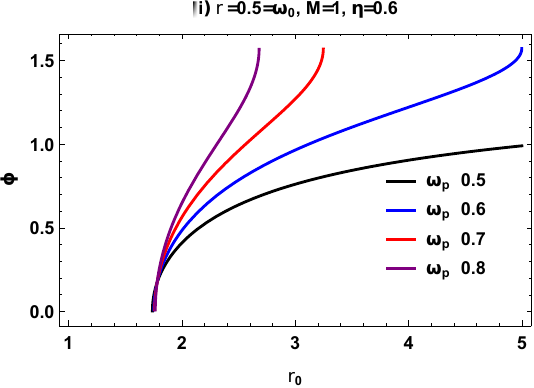}\includegraphics[width=7cm]{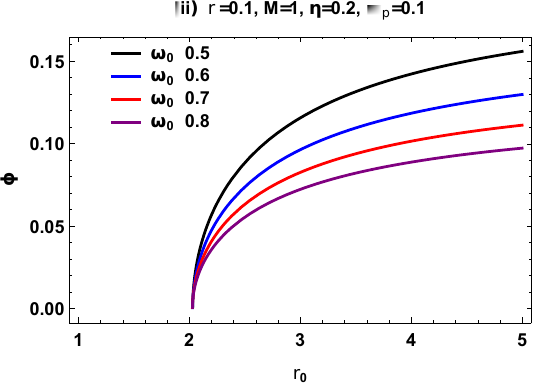}\\
{Figure 4: The plasma shadow $\Phi$ versus  radius $r_{0}$ for various $\omega_{p}$, $\omega_{0}$ and fixed $\tilde{r}=0.5=\omega_{0}, M=1~\text{and}~ \eta=0.6$ in the left plot while for fixed $\tilde{r}=0.1=\omega_{p}, M=1~\text{and}~\eta=0.2$ in the right plot.}
\end{center}
The left plot of {\bf Fig. 4} shows the behavior of the angular radius of shadow with static observer at distinct positions $r_{0}$ for varying $\omega_{p}$ in low density medium. The $\Phi$ continuously goes on increasing but the radius of shadow becomes smaller. The plasma shadow rises for rising values of $\omega_{p}$. In the right plot of {\bf Fig. 4}, the $\Phi$  exponentially rises for rising $r_{0}$. It is also observable that the shadow radius in low density plasma is smaller as compared to {\bf Fig. 3}. The {\bf Fig. 4} states that the BH shadow goes on increasing for different variations of distinct observer $r_{0}$. To get a very large shadow radius in strong gravity, the frequency parameter must be extremely small.

\section{Conclusions}
In this research,  we have examined deformation parameter effect on metric function by considering the methodology of Keeton and Petters and the framework ofÂ PPNÂ is a direct way for dealing with all types of gravity theories in which the weak-deflection limit is stated as a series expansion withÂ single variable $M$. The deflection angle depends on the deformation parameter ($\eta$), the impact parameter ($b$) and the mass ($M$). The results we have obtained for BH solutions in the presence of non-plasma mediums show that deflection angle has a direct relationship with mass $M$ and a parameter $\eta$, indicating that BH with greater mass has greater gravitational pull and bends the light passing by it at large angles. While BH with smaller mass deflect the light at a smaller angle. Additionally, we observe that deflection angle has an inverse relationship with impact parameter $b$, demonstrating that a smaller value of the impact parameter results in a larger deflection angle and vice versa. Moreover, we conclude that for $\eta=0$, we recover the deflection angle for Schwarzschild BH.

We have examined the BH shadow radius in plasma as well as low density plasma medium by using the Hamilton equation to describe photon motion. To do so, by using a new ray-tracing algorithm, we numerically solved the Hamilton equations to determine the BH shadow radius. The size and shape of the BH shadow can alter when there is plasma surrounding a BH because the trajectory of a photon is dependent on the photon frequency. The photon-absorbing and scattering actions of plasma electrons are not taken into consideration. Further, we ignore the plasma particle's gravitational field. Furthermore, the size of shadow always shrinks when plasma is present and for $\eta=0$ and $r_{ph}=3M$ in Eq. (\ref{b37}), we get the radius of shadow for Schwarzschild BH. At low frequency plasma density, the shadow of a BH is shown to appear on the observer's screen. The BH shadow is identical to that BH at high frequency, and the influence of the plasma is minimal. Furthermore, the presence of a plasma may result in the formation of a number of structures in the BH image.

Moreover, we have graphically analyzed the behavior of shadow for distinct positions under the effects of plasma frequency $w_{p}$ and $w_{0}$ in plasma as well as low density plasma medium. We have observed that $\Phi$ slowly decreases and then, it gets a flat form for rising $r_{0}$. It is notable that the $\Phi$ rises for rising values of $\omega_{p}$.
For changing values of $\omega_{0}$, the $\Phi$ slowly decreases till $r_{0}\rightarrow \infty$. The $\Phi$ decreases with rising $\omega_{0}$.
The BH shadow is smaller for different variations of $r_{0}$. To get large shadow radius in strong gravity, the frequency parameter must be extremely small. The $\Phi$ reduces for rising values of $\omega_{p}$ in the region $1\leq r_{ph}\leq5$. It shows a rising attitude for rising $r_{ph}$ as well as the radius of shadow increases for rising values of $\omega_{0}$. In low density medium, the $\Phi$ continuously goes on increasing but the radius of shadow becomes smaller. The plasma shadow rises for rising values of $\omega_{p}$. The $\Phi$  exponentially rises for rising $r_{0}$ for varying $\omega_{0}$. It is also observable that the shadow radius in low density plasma is smaller as compared to {\bf Fig. 3}. The BH shadow is goes on increasing for different variations of distinct observer $r_{0}$. To get a very large shadow radius in strong gravity, the frequency parameter must be extremely small. We have simply taken into account the photon motion in a stationary, non-magnetized, pressure-free plasma in the background of a BH here as a first step towards understanding the influence of ionised matter on the various astrophysical compact objects. Future research should naturally build on these findings to incorporate more complex yet realistic plasma models.

We anticipate that the observation will provide us with important hints and surprises regarding space-time and the distribution of matter in the vicinity of the BH, as revealed by the future results of the Event Horizon Telescope (EHT). However, expecting that EHT can effectively limit the plasma density distribution's properties is suggested. Although the pulling effect extends the shadow of a revolving BH in this position.

\end{document}